\documentclass[aps,prl,reprint,superscriptaddress]{revtex4-1}

\usepackage{graphicx}
\usepackage{epsfig}
\usepackage{amsmath,bbm,amssymb,bm,times}
\usepackage[colorlinks,linkcolor=blue,citecolor=blue]{hyperref}

\usepackage{stmaryrd}
\usepackage[normalem]{ulem}
\usepackage{comment}
\usepackage[usenames,dvipsnames]{color}
  
\definecolor{dblue}{rgb}{0.0, 0.0, 0.6}

\definecolor{dgreen}{rgb}{0.0, 0.5, 0.0}

\newcommand{\eqn}[1]{
\begin{eqnarray}
	#1
\end{eqnarray}
}

\begin{document}

\title{Topological-to-Topological Transition Induced by On-Site Nonlinearity in a One-Dimensional Topological Insulator}
\author{Kazuki Sone}
\email{sone@rhodia.tsukuba.ac.jp}
\affiliation{Department of Physics, University of Tsukuba, Tsukuba, Ibaraki 305-8571, Japan}
\author{Yasuhiro Hatsugai} 
\affiliation{Department of Physics, University of Tsukuba, Tsukuba, Ibaraki 305-8571, Japan}
\date{\today}

\begin{abstract}
Recent studies have extended the notion of band topology to nonlinear systems by defining nonlinear counterparts of eigenvalue problems. They have found the nonlinearity-induced topological transition, while it has required complicated nonlinearity such as off-diagonal one. Thus, the existence of nonlinearity-induced transitions has been unclear under homogeneous on-site nonlinearity, which is ubiquitously found in nature. We here reveal that such on-site nonlinearity can induce transitions of topological modes, where topological modes converging to zero begin to converge to nonzero values. Since such nonlinearity-induced transition remains the bulk band topology unchanged, we can regard it as a transition from a conventional topological mode to one unique to nonlinear systems. We analyze a nonlinear eigenvalue problem by rewriting it to a dynamical system in the spatial direction and clarify that the nonlinearity-induced transition is a result of the bifurcation in the spatial dynamics. We also propose a possible setup to observe the nonlinearity-induced transition that uses a gradual amplification of nonlinear waves. These results provide a general designing principle of topological insulators controlled by nonlinearity.

\end{abstract}

\maketitle

Nonlinear effects appear in various physical systems, including photonics \cite{Berger1998}, fluids \cite{Korteweg1895}, and interacting bosons \cite{Khaykovich2002}. In particular, on-site nonlinearity lies at the center of studies of nonlinear physics, because it is obtained from the mean-field approximation of short-range many-body interactions. The leading order of such on-site nonlinearity is known as the Kerr nonlinearity \cite{Boyd2003} in photonics and also appears in the Gross-Pitaevskii equation \cite{Gross1961,Pitaevskii1961} of cold atoms. The on-site nonlinearity induces unique wave phenomena such as solitons \cite{Shabat1972,Morandotti1999,Kartashov2011}.

Recent studies have also investigated the effect of nonlinearity on band topology. Materials with nontrivial band topology, such as topological insulators \cite{Hasan2010,Qi2011}, exhibit topological gapless modes localized at the edge of the sample, which is known as the bulk-boundary correspondence \cite{Laughlin1981,Thouless1982,Hatsugai1993,Ryu2002,Hatsugai2009}. While the notion of band topology can be extended to dispersion relations in various classical and quantum systems \cite{Haldane2008,Prodan2009,Hafezi2011,Hafezi2013,Atala2013,Jotzu2014,Kane2014,Yang2015,Delplace2017,Souslov2017,Shankar2017}, the nonlinear effect in such systems can alter the bulk-boundary correspondence \cite{Lumer2013,Chen2014,Leykam2016,Harari2018,Smirnova2019,Zangeneh2019,Darabi2019,Zhang2020,Maczewsky2020,Smirnova2020,Ota2020,Ivanov2020,Lo2021,Mukherjee2021,Kotwal2021,Sone2021,Wachtler2023,Jurgensen2021,Fu2022,Mostaan2022,Li2022,Ezawa2021,Hadad2016,Mochizuki2021}. By extending the eigenvalue problem to nonlinear systems, some previous studies \cite{Tuloup2020,Zhou2022,Sone2024a,Sone2024b,Zhou2024} have defined the nonlinear topological invariants and discussed their bulk-boundary correspondence. They have also revealed that nonlinear topological invariants and corresponding gapless modes depend on the amplitude of nonlinear waves, which is termed the nonlinearity-induced topological transition. However, models of the nonlinearity-induced topological transitions have utilized complicated nonlinear terms such as off-diagonal ones, and thus it was unclear whether such transitions can occur by on-site nonlinearity that 
ubiquitously appears in experimental setups. From the theoretical viewpoint, since the on-site nonlinearity breaks the sublattice symmetry assumed in previous research \cite{Sone2024b,Zhou2024}, one needs to extend the analytical technique on nonlinear topological modes under the absence of such symmetries.

In this Letter, we reveal the existence of the transition of topological localized modes induced by uniform on-site nonlinearity. In this nonlinearity-induced transition, the long-range behavior of topological modes is changed; after the transition, they converge to nonzero even in the thermodynamic limit. Such localized modes with nonzero remaining amplitudes are similar to those observed in the nonlinearity-induced topological transition in previous studies. However, the nonlinearity-induced transition discussed here accompanies no gap closing and thus remains the band topology unchanged. 
We analyze such nonlinearity-induced transitions by using a dynamical system in the spatial direction, which is a nonlinear extension of a transfer matrix. We find the bifurcation of fixed points in the spatial dynamics and that those bifurcations correspond to the nonlinearity-induced transition of topological localized modes. We also propose the observation protocol of the nonlinearity-induced transition by the gradual amplification of nonlinear waves. These results clarify the universal correspondence between the transition of topological modes and the bifurcation in dynamical systems. In addition, because of the feasibility of on-site nonlinearity, our proposal is directly relevant to the experimental realization of nonlinearity-induced transitions of topological modes.

\begin{figure}[t]
  \includegraphics[width=70mm,bb=0 0 550 271,clip]{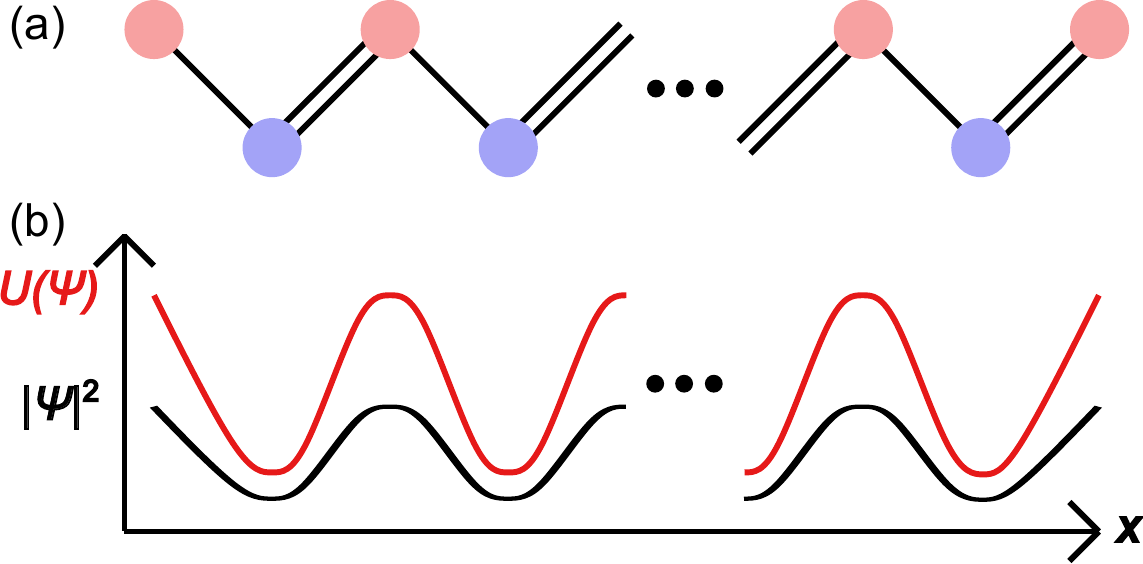}
  \caption{\label{fig1} Schematic of the nonlinear SSH model. (a) The nonlinear SSH model and the boundary condition. The model has two sublattices represented by the red and blue circles. The strengths of linear hoppings are alternately modulated. The double lines represent stronger hopping than the single lines. Sublattice A (red circles) lies at both open boundaries. Under this boundary condition, we obtain a localized mode at either end. (b) Effective potential induced by the on-site nonlinearity. The red curve shows the effective on-site potential, which is proportional to the square of the nonlinear wave (the black curve).
  }
  \end{figure}

{\it Setup.---} 
We consider the nonlinear Su-Schrieffer-Heeger (SSH) model \cite{Su1979} with on-site Kerr nonlinearity \cite{Chen2014,Hadad2016,Engelhardt2017,Dobrykh2018,Wang2019,Xia2020,Guo2020,Ezawa2021,Chaunsali2021,Ma2021,Zhou2022} shown in Fig.~\ref{fig1}. Its time evolution is described as
\begin{eqnarray}
&{}& \!\!\!\! i\partial_t \Psi_A(x) \!  = \! a\Psi_B(x) + b\Psi_B(x-1) + U(\Psi_A(x)) \Psi_A(x), \ \ \ \ \ \label{SSH1} \\
&{}& \!\!\!\! i\partial_t \Psi_B(x) \!  = \! a\Psi_A(x) + b\Psi_A(x+1) + U(\Psi_B(x)) \Psi_B(x), \ \ \ \ \ \label{SSH2} \\
&{}& \!\!\!\! U(\Psi_{A,B}(x)) \! = \! \kappa|\Psi_{A,B}(x)|^2 \label{onsite_nonlinearity},
\end{eqnarray}
where $U(\Psi_{A,B}(x))$ corresponds to the self-induced potential and determines the strength of the Kerr nonlinearity at each site. By assuming a periodic solution $\Psi_{A,B}(x;t)=e^{-iEt}\Psi_{A,B}(x)$, we obtain the nonlinear eigenvalue problem 
\begin{eqnarray}
&{}& \!\! E \Psi_A(x) \!  = \! a\Psi_B(x) + b\Psi_B(x-1) + U(\Psi_A(x)) \Psi_A(x), \ \ \ \ \ \label{SSH1_eigen} \\
&{}& \!\! E \Psi_B(x) \!  = \! a\Psi_A(x) + b\Psi_A(x+1) + U(\Psi_B(x)) \Psi_B(x). \ \ \ \ \ \label{SSH2_eigen}
\end{eqnarray}
We further assume spatially periodic solutions $\Psi_{A,B}(x) = e^{ikx}\psi_{A,B}(k)$ termed the Bloch ansatz \cite{Sone2024a,Sone2024b} and finally obtain
\begin{equation}
  \left(
  \begin{array}{cc}
   \kappa|\psi_A|^2 & a+be^{-ik} \\
   a+be^{ik} & \kappa|\psi_B|^2
  \end{array}
  \right)
  \left(
  \begin{array}{c}
   \psi_A\\
   \psi_B
  \end{array}
  \right) = E  \left(
  \begin{array}{c}
   \psi_A\\
   \psi_B
  \end{array}
  \right), \label{ssh_eigeneq}
\end{equation}
which is a wavenumber-space description of the nonlinear eigenvalue problem and corresponds to the Bloch Hamiltonian in linear systems.

Due to the absence of the superposition law, nonlinear eigenvectors depend on their norm. We note that scaling of the wavefunction $\psi \rightarrow \lambda \psi$ induces the change of the strength of nonlinearity $\kappa \rightarrow \lambda^2 \kappa$. In the following, we fix $\kappa$ to be $\kappa=1$ and instead modify the strength of nonlinearity by considering eigenvectors with different norms.
Specifically, in the calculation of bulk solutions, we focus on special solutions whose amplitudes are constant independently of the wavenumber, $|\psi_{A}(k)|^2+|\psi_{B}(k)|^2=w$ \cite{Sone2024a,Sone2024b}.

By analytically solving Eq.~\eqref{ssh_eigeneq}, we obtain two types of bulk eigenvectors; ones obtained in the linear limit 
\begin{equation}
(\psi_A(k),\psi_B(k)) = \sqrt{w/2}(1, \pm e^{i\theta}) \label{bulk_es1}
\end{equation}
with the eigenvalues being $E=|a+be^{ik}|+\kappa w / 2$ and $\theta$ representing $\theta = \arg(a+be^{ik})$, and ones without linear counterparts 
\begin{equation}
(\psi_A(k),\psi_B(k)) = (r_{\pm},  e^{i\theta}|a+be^{ik}|/(\kappa r_{\pm})) \label{bulk_es2}
\end{equation}
with the eigenvalues being $E=\kappa w$ and $r_{\pm}$ and $\theta$ representing $r_{\pm}=\sqrt{\left[\kappa w \pm \sqrt{(\kappa w)^2 - 4|a+be^{ik}|^2} \right] / (2\kappa)}$ and $\theta = \arg(a+be^{ik})$. The latter type of bulk eigenvectors appears only at large amplitude $w>2|a+be^{ik}|/\kappa$. By using these bulk nonlinear eigenvectors, we can define the nonlinear Berry phase \cite{Zhou2022,Zhou2024} as
\begin{equation}
\nu(w) = \frac{1}{w} \int_0^{2\pi} dk \left[\psi_A(k)^{\ast} \partial_k \psi_A(k) + \psi_B(k)^{\ast} \partial_k \psi_B(k)\right].
\end{equation}
Since the nonlinear SSH model has the time-reversal and spatial inversion symmetries \cite{supple}, this nonlinear Berry phase is quantized, $\nu(w) = 0,\,\pi\ ({\rm mod}\ 2\pi)$, and used as a topological invariant.

One can formulate the bulk-boundary correspondence of the nonlinear system as follows \cite{Sone2024a,Sone2024b}: one considers the open boundary condition with a fixed edge amplitude, $|\Psi_A(1)|^2+|\Psi_B(1)|^2 = w$ and $\Psi_B(0)=0$ with $w$ being that used to calculate the nonlinear Berry phase. Then, the nonzero (zero) Berry phase corresponds to the existence (absence) of localized gapless modes. While this bulk-boundary correspondence looks similar to the linear one, nonlinearity can produce anti-localized gapless modes in trivial systems with $\nu(w)=0\ ({\rm mod}\ 2\pi)$. However, the existence of anti-localized modes is consistent with the above bulk-boundary correspondence because it focuses on the existence or absence of localized modes. It is also noteworthy that nonlinear topological modes can converge to nonzero values in the limit of $x\rightarrow\infty$, which we cannot normalize and thus is not regarded as conventional eigenvectors in linear cases. In the nonlinear SSH model \eqref{ssh_eigeneq}, we can numerically confirm this bulk-boundary correspondence in a wide range of the strength of nonlinearity, while we find that the on-site nonlinearity induces the transition of topological modes from decaying modes to extended modes.

\begin{figure}[t]
  \includegraphics[width=86mm,bb=0 0 582 220,clip]{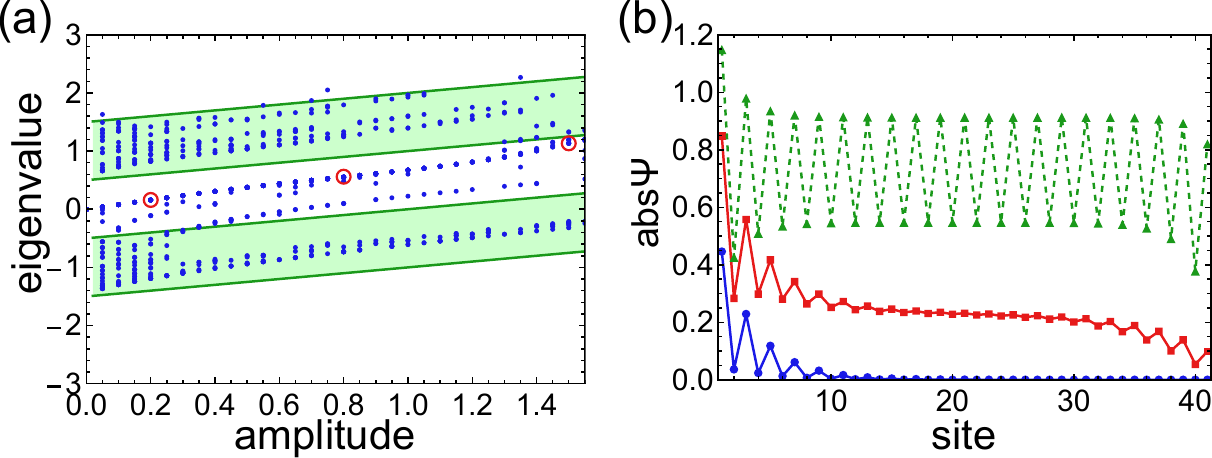}
  \caption{\label{fig2} Eigenvalues and eigenvectors at each amplitude. (a) Amplitude dependence of the nonlinear eigenvalue. Blue dots represent the nonlinear eigenvalues at each edge amplitude $|\Psi_A(1)|^2+|\Psi_B(1)|^2 = w$ numerically obtained under the open boundary condition. The green bands show the analytically obtained bulk bands. (b) Nonlinear topological modes and their nonlinearity-induced transition. The blue, red, and green dashed polylines show the spatial distribution of absolute values of the topological modes, whose edge amplitudes are $w=0.2$, $w=0.8$, and $w=1.5$, respectively. Their eigenvalues are presented in panel (a) by the red circles. If the edge amplitude is small (blue polyline), the topological mode converges to zero in the bulk, while above a critical amplitude, nonzero amplitudes remain in the bulk. If we consider larger amplitudes, the dimerization is broken, and two sublattices have different amplitudes as shown by the green dashed polyline. 
  }
  \end{figure}

{\it Nonlinearity-induced transition of edge modes.---} 
By numerically solving the nonlinear eigenvalue problem [\eqref{SSH1_eigen} and \eqref{SSH2_eigen}], 
we find the transition in the converging behavior of the topological edge modes at a critical amplitude. To solve the nonlinear eigenvalue problem, we use the quasi-Newton method \cite{Sone2024a,Broyden1965}. We consider the open boundary condition in Fig.~\ref{fig1}(a), where we cut the intercell (intracell) bonding at the left (right) edge. Under such a boundary condition, $\nu(w)=\pi\ ({\rm mod}\ 2\pi)$ corresponds to the appearance of edge modes at the left end, while even in the case of $\nu(w)=0\ ({\rm mod}\ 2\pi)$, we obtain edge modes at the right end \cite{Kane2014}. The advantage of assuming this boundary condition is that we obtain exact solutions of edge modes without considering superpositions of left- and right-localized modes and thus can suppress the effect of the boundary condition at the edge opposite to the direction of the localization. In the following, we fix the parameters $a=-0.5$ and $b=1$ \footnote{The qualitative behaviors (cf. the nonlinearity-induced transition of edge modes) are independent of the parameters as long as the nonlinear Berry phase is nonzero, while the signs in transition points can be changed.}. where the nonlinear Berry phase becomes $\nu(w)=\pi\ ({\rm mod}\ 2\pi)$ at any $w$, and thus left-localized modes appear.

Figure \ref{fig2}(a) shows the amplitude dependence of the nonlinear eigenvalues. Since the effective on-site potential $U(\Psi)$ is proportional to the amplitude $w$, the nonlinear eigenvalues are shifted following the change of the amplitude, which is consistent with the analytical bulk solutions. We also confirm the existence of gapless modes. Such gapless modes are localized at the left edge, i.e., $|\Psi_A(x_{\rm bulk})|^2+|\Psi_B(x_{\rm bulk})|^2 < |\Psi_A(1)|^2+|\Psi_B(1)|^2$ at the bulk site $x_{\rm bulk}\gg 1$. However, we also find that the edge modes have nonzero amplitudes at the bulk sites if the edge amplitude $|\Psi_A(1)|^2+|\Psi_B(1)|^2$ is larger than a critical value $w_c \sim 0.6$ (cf. Fig.~\ref{fig2}(b)), where the eigenvalue of the edge mode is also larger than that of bulk modes in the linear limit, $E_{\rm edge} > a+b$. This amplitude dependence implies the nonlinearity-induced transition of topological modes from decaying localized modes to nonvanishing ones in the semi-infinite system. We note that remaining amplitudes in the limit of $x\rightarrow\infty$ can also be seen in edge modes emerging by nonlinearity-induced topological phase transitions discussed in previous papers \cite{Hadad2016,Zhou2022,Sone2024a,Sone2024b}. Unlike them, the transition observed in Eqs.~\eqref{SSH1_eigen} and \eqref{SSH2_eigen} accompanies no gap closing, and thus the bulk topology is unchanged, while it is still induced by the change of the strength of the nonlinearity $U(\Psi_{A,B}(x))$.

\begin{figure}[t]
  \includegraphics[width=60mm,bb=0 0 305 350,clip]{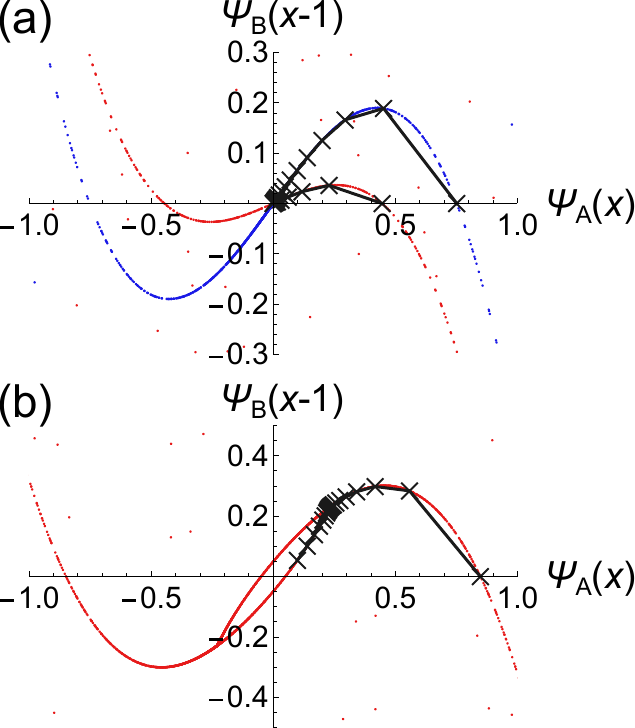}
  \caption{\label{fig3} Stable manifolds and bifurcation of the dynamical system in the spatial direction. (a) Stable manifolds at different eigenvalues. The red and blue dots show the stable manifolds corresponding to edge modes whose edge amplitudes are $w=0.2$ and $0.5$, respectively. Due to the limitation of the numerical technique, there appear dots far from the curves, which are irrelevant to the stable manifold. The black crosses represent $(\Psi_A(x),\Psi_B(x-1))$ of the edge modes at each $x$. We can confirm that each component of the edge mode lies along the stable manifolds, and thus expect that the stable manifold determines the behavior of edge modes. (b) Stable manifolds after the transition. The red dots show the stable manifold corresponding to an edge mode whose edge amplitude is $w=0.8$. The black crosses represent $(\Psi_A(x),\Psi_B(x-1))$ of the edge mode at each $x$. Nonzero fixed points $\Psi_A(x)=\Psi_B(x)\simeq \pm 0.23$ appear in this parameter regime. As in panel (a), the components of the edge mode lie along the stable manifolds. 
   }
  \end{figure}

{\it Dynamical system in the spatial direction.---} 
To clarify the existence of the nonlinearity-induced transition of topological in-gap states under the semi-infinite boundary condition, we analyze the spatial distribution of topological modes by rewriting the nonlinear eigenvalue problem into a discrete dynamical system in the spatial direction. By assuming the nonlinear eigenvalue $E$ as a parameter, we obtain the following nonlinear dynamical system from Eqs.~\eqref{SSH1_eigen} and \eqref{SSH2_eigen}:
\begin{eqnarray}
&{}& \!\!\!\!\!\!\! \Psi_A(x+1) \! = \! \frac{-a\Psi_A(x)-\kappa|\Psi_B(x)|^2\Psi_B(x)+E\Psi_B(x)}{b}, \ \ \ \ \ \ \label{transfer1}\\
&{}& \!\!\!\!\!\!\! \Psi_B(x+1) \nonumber\\
&{}& \!\!\!\!\!\!\! = \! \frac{-b\Psi_B(x)-\kappa|\Psi_A(x+1)|^2\Psi_A(x+1)+E\Psi_A(x+1)}{a},\ \ \ \ \ \ \ \label{transfer2}
\end{eqnarray}
which corresponds to a nonlinear counterpart of the transfer-matrix representation of an eigenequation. 
Under the open boundary condition at the left edge $\Psi_B(0)=0$, we can obtain the spatial distribution of a nonlinear eigenvector in the right semi-infinite system by iteratively calculating the above equations. We note that due to the time-reversal symmetry of the system, all the parameters in Eqs.~\eqref{transfer1} and \eqref{transfer2} are real, and we can assume that topological modes also have real eigenvectors.

The dynamical system [\eqref{transfer1} and \eqref{transfer2}] has some fixed points and periodic solutions. Except for the trivial fixed point $(\Psi_A(x),\Psi_B(x))=(0,0)$, those fixed points and periodic solutions correspond to bulk eigenvectors [\eqref{bulk_es1} and \eqref{bulk_es2}] obtained from the Bloch ansatz, because periodicity in Eqs.~\eqref{transfer1} and \eqref{transfer2} indicates the spatial periodicity of the nonlinear eigenvector.

From the viewpoint of the dynamical system, we can regard the converging solutions from far points to fixed points or periodic solutions as localized modes because the corresponding nonlinear eigenvectors have larger amplitudes at the edge, $x=1$, than in the bulk, $x\gg 1$. As demonstrated in Fig.~\ref{fig3}(a), orbits of such converging solutions lie on the stable manifold of a fixed point $(A_0,B_0)$, which is defined as a point set of $ (\Psi_A,\Psi_B)$ such that $(\lim_{x\rightarrow\infty}\Psi_A(x),\lim_{x\rightarrow\infty}\Psi_B(x)) = (A_0,B_0) $ when the initial condition is $(\Psi_A(0),\Psi_B(0)) = (\Psi_A,\Psi_B)$. Fig.~\ref{fig3}(a) also shows the amplitude dependence of the stable manifold of the fixed point at $(0,0)$. We can confirm that the stable manifold crosses the $\Psi_B=0$ axis at a larger point as the nonlinear eigenvalue $E$ becomes larger. Thus, the topological localized mode that satisfies the boundary condition $\Psi_B(0)=0$ has a larger eigenvalue at a larger amplitude as is seen in Fig.~\ref{fig2}(a). It is noteworthy that the dynamical system [\eqref{transfer1} and \eqref{transfer2}] is discrete, while we still obtain curve-shaped stable manifolds at least in the range where we can numerically estimate their shapes \cite{supple}. 

The converging solutions exhibit a transition at $E=a+b$ due to the bifurcation of the fixed point at $(0,0)$. Corresponding to the emergence of bulk modes at that eigenvalue, the stability of the fixed point $(0,0)$ is changed; it is a saddle point at $E<a+b$ and becomes a center (i.e., linear stability is marginal) at $E>a+b$ \cite{supple}. Instead, we obtain steady solutions $(\Psi_A(x),\Psi_B(x)) = (\pm \sqrt{w/2}, \pm \sqrt{w/2})$, who has a one-dimensional stable manifold shown in Fig.~\ref{fig3}(b). Therefore, the solutions corresponding to topological localized modes converge to the nonzero steady solutions $(\Psi_A(x),\Psi_B(x)) =  (\pm\sqrt{w/2}, \pm \sqrt{w/2})$ if the nonlinear eigenvalue is larger than $a+b$. Since the eigenvalue becomes larger as the amplitude does, such a transition must occur at a certain amplitude, which indicates the existence of the nonlinearity-induced transition of topological localized modes.

{\it Further transition at a larger amplitude.---} 
Corresponding to the emergence of the additional bulk modes \eqref{bulk_es2}, the steady solutions $(\Psi_A(x),\Psi_B(x)) = (\pm\sqrt{w/2}, \pm \sqrt{w/2})$ of the dynamical system [\eqref{transfer1} and \eqref{transfer2}] also exhibit a bifurcation. Since such a bifurcation changes the linear stability of the periodic solution and convergent values of the nonlinear dynamics, we can observe a further nonlinearity-induced transition of topological localized modes by considering larger edge amplitudes than $w_c'\sim 1.4$, at which the nonlinear eigenvalue of the localized mode becomes $E=2(a+b)$. In fact, the localized mode represented by the green dashed polyline in Fig.~\ref{fig2}(b) shows different absolute values between A and B sublattices, unlike that obtained at smaller edge amplitudes and corresponding smaller eigenvalues, $a+b<E<2(a+b)$. As in the nonlinearity-induced transition at $E=a+b$, we can analytically confirm the change of the linear stability of the periodic solution $(\Psi_A(x),\Psi_B(x)) = (\pm\sqrt{w/2}, \pm \sqrt{w/2})$ \cite{supple}. Thus, this further nonlinearity-induced transition is also understood as the change of the convergent value associated with the bifurcation of the periodic solution.

\begin{figure}[t]
  \includegraphics[width=86mm,bb=0 15 600 218,clip]{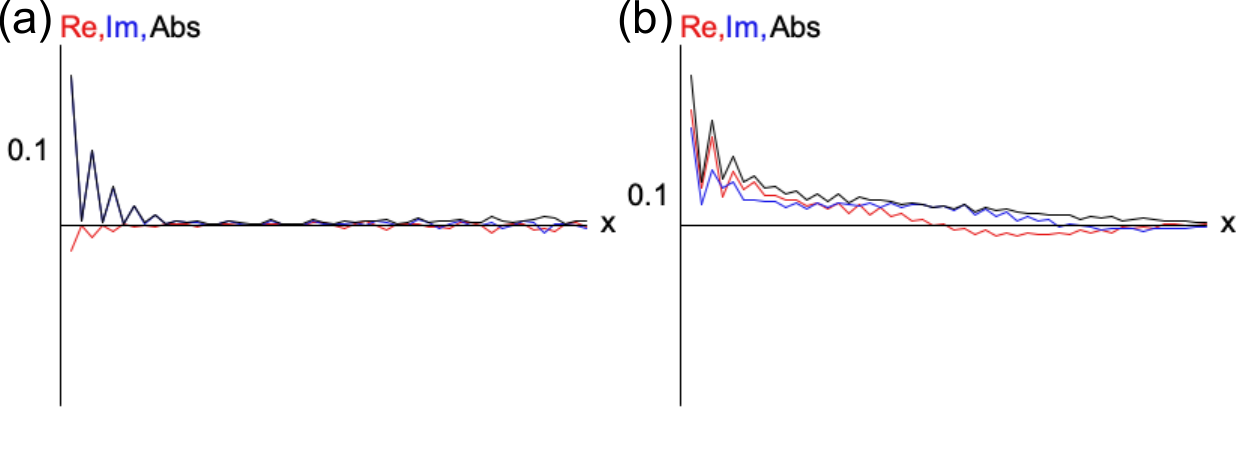}
  \caption{\label{fig4} Dynamical demonstration of the nonlinearity-induced transition. (a) Localized waves observed before introducing the gain. The red, blue, and black polylines show the real, imaginary parts, and absolute values of the nonlinear wave, respectively. The localized wave decays to zero in the bulk. (b) Localized waves after the amplification. We can confirm the nonzero amplitudes and dimerization in the bulk, which are consistent with the nonlinear eigenvector obtained in Fig.~\ref{fig2}.
  }
  \end{figure}

{\it Observation protocol of the nonlinearity-induced transition by gradual amplification.---} 
We here propose a possible observation protocol of the nonlinearity-induced transition by the continuum deformation of nonlinearity. As we have discussed above, the change of the wave amplitude $w$ also alters the nonlinear self-induced potential $U(\Psi_{A,B}(x))=\kappa |\Psi_{A,B}(x)|^2$. Thus, we can continuously tune the strength of nonlinearity by gradually amplifying the nonlinear wave, which leads to possible observation of the nonlinearity-induced transition. The gradual amplification of the nonlinear wave is realized by simply introducing a uniform gain. We note that since the gain and the on-site Kerr nonlinearity are ubiquitous in, e.g., photonics, one can realize the nonlinear SSH model and the observation protocol in existing setups of topological photonics \cite{Dobrykh2018,Guo2020,StJean2017,Parto2018,Zhao2018}.

We numerically simulate the nonlinear dynamics \eqref{SSH1}, \eqref{SSH2} under the existence of a temporally controlled gain (cf. Fig.~\ref{fig4}). We consider the system size $L=1000$. As an initial condition, we consider the excitation only at the edge site $\Psi_A(1) = \sqrt{0.1}$, $\Psi_A(x)=0$ ($x\geq2$), $\Psi_B(x)=0$ and wait until $t=t_0=200$, when the bulk-propagating wave goes far from the left edge. Then, we introduce a homogeneous gain whose strength depends on the time as $c(t) = \left[ 2(t-t_0)+1/c_0 \right]^{-1}$. The time dependence of $c(t)$ is necessary to suppress the exponential divergence of the strength of nonlinearity. Finally, we switch off the gain at $t=300$. In the numerical result, we can confirm that the localized wave penetrates into the bulk at a sufficiently large amplitude. Since the remaining amplitude is characteristic of topological modes after the nonlinearity-induced transition, this penetration implies the existence of the nonlinearity-induced transition.

{\it Discussion.---} 
We revealed the existence of the nonlinearity-induced transition in the nonlinear SSH model with on-site nonlinearity. Before the transition, topological localized modes decay to zero like those in linear cases, while after the transition, they converge to nonzero values in the limit of $x\rightarrow\infty$. To analyze the topological modes in the nonlinear SSH model, we rewrite the nonlinear eigenvalue problem into a dynamical system in the spatial direction, which is a nonlinear generalization of a transfer matrix. Then, the nonlinearity-induced transition is understood as a bifurcation of a fixed point at the origin. We note that the nonlinearity-induced transition can occur in a wide range of nonlinear systems because the appearance of bulk bands must accompany the bifurcations of fixed points in the spatial dynamical system.

From an experimental viewpoint, one can realize the nonlinearity-induced transition by gradual amplification of a nonlinear localized mode, which introduces the continuum deformation of nonlinearity. We note that nonlinear SSH models have been discussed in e.g. photonics \cite{Dobrykh2018,Guo2020}. In realistic setups, however, higher-order nonlinearity than the third-order one considered in this work can affect the nonlinear dynamics. Thus, investigating the effect of such higher-order terms remains a crucial issue to experimentally realize the nonlinear topological phenomena.

While a previous study \cite{Sone2024b} has analyzed the bulk-boundary correspondence in nonlinear systems by using a dynamical system in the spatial direction, it has assumed the sublattice symmetry. The present work has extended such a technique to the system without the sublattice symmetry. Since without the sublattice symmetry, topological modes can have nonzero eigenvalues, we need to consider a dynamical system where the nonlinear eigenvalue plays the role of an additional parameter. However, we can still relate the topological localized modes with the converging solutions to fixed points or periodic solutions (in other words, their stable manifolds), and discuss the nonlinearity-induced transitions in view of their bifurcations. The analytical technique without fixing the nonlinear eigenvalue is also applicable to higher-dimensional systems with arbitrary symmetries and thus should be useful to investigate the bulk-boundary correspondence in general nonlinear systems.

We thank Natsuko Ishida and Satoshi Iwamoto for valuable discussions. K.S. is supported by JSPS KAKENHI Grant Number JP24K22848. Y.H. is supported by JSPS KAKENHI Grant Number JP23K25788 and JST CREST, Grant No. JPMJCR19T1.

\bibliography{reference}

\widetext
\pagebreak
\begin{center}
\textbf{\large Supplementary Materials}
\end{center}

\renewcommand{\theequation}{S\arabic{equation}}
\renewcommand{\thefigure}{S\arabic{figure}}
\renewcommand{\bibnumfmt}[1]{[S#1]}
\setcounter{equation}{0}
\setcounter{figure}{0}

\subsection{\label{sec1}Time-reversal and space-inversion symmetries of the nonlinear Su-Schrieffer-Heeger (SSH) model.}
One can define symmetries in nonlinear systems by using a composition of functions \cite{Sone2024b}. Specifically, when the nonlinear eigenvalue problem is described as ${\bf f}(\boldsymbol{\Psi}) = E\boldsymbol{\Psi}$, ${\bf f}(\boldsymbol{\Psi})=(f_{1,1}(\boldsymbol{\Psi}),\ldots,f_{N,L}(\boldsymbol{\Psi}))$ with $N$ and $L$ being the internal degree of freedom and the system size and $\boldsymbol{\Psi}$ being the vector $\boldsymbol{\Psi}=(\Psi_1(1),\ldots,\Psi_N(L))$, the time-reversal and space-inversion symmetries are defined as
\eqn{
T^{-1} {\bf f}(T\boldsymbol{\Psi}) &=& {\bf f}^{\ast}(\boldsymbol{\Psi}) \ \ \text{(time-reversal symmetry),}\\
P^{-1} {\bf f}(P\boldsymbol{\Psi}) &=& {\bf f}(\boldsymbol{\Psi}) \ \ \text{(space-inversion symmetry),}
}
where $T$ and $P$ are time-reversal and space-inversion linear operators, respectively. Assuming the Bloch ansatz $\Psi_j(x)=e^{ikx}\psi_j$, one can extend these definitions of symmetries to the wavenumber-space description of the nonlinear eigenvalue problem ${\bf f}(\boldsymbol{\psi} ;k) = E(k)\boldsymbol{\psi}$, $\boldsymbol{\psi}=(\psi_1,\ldots,\psi_N)$ as
\eqn{
T^{-1} {\bf f}(T\boldsymbol{\psi} ; k) &=& {\bf f}^{\ast}(\boldsymbol{\psi} ; -k) \ \ \text{(time-reversal symmetry),}\\
P^{-1} {\bf f}(P\boldsymbol{\psi} ; k) &=& {\bf f}(\boldsymbol{\psi} ; -k) \ \ \text{(space-inversion symmetry).}
}
We note that under these symmetries, the nonlinear band structure $E(k)$ has the same symmetry as in linear cases; for example, if the nonlinear system has the time-reversal symmetry and all the nonlinear eigenvalues are real, the set of nonlinear eigenvalues at $k$, $\{E(k)\}$, is the same as $\{E(-k)\}$.

The nonlinear SSH model in Eqs.~(1-6) in the main text has the time-reversal and space-inversion symmetries. Here, we define the time-reversal and space-inversion operators as $T=I $ and $P=\sigma_x$ with $I$ being the $2\times2$ identity matrix and $\sigma_i$ being the Pauli matrix. Then, if we rewrite the left-hand side of the nonlinear eigenequation in Eq.~(6) in the main text as
\eqn{
 {\bf f}(\boldsymbol{\psi};k) =   \left(
  \begin{array}{cc}
   \kappa|\psi_A|^2 & a+be^{-ik} \\
   a+be^{ik} & \kappa|\psi_B|^2
  \end{array}
  \right)
  \left(
  \begin{array}{c}
   \psi_A\\
   \psi_B
  \end{array}
  \right),
}
we can confirm the time-reversal and space-inversion symmetries by following calculations:
\eqn{
T^{-1}{\bf f}(T\boldsymbol{\psi};k) &=& \left(
  \begin{array}{cc}
   \kappa|\psi_A|^2 & a+be^{-ik} \\
   a+be^{ik} & \kappa|\psi_B|^2
  \end{array}
  \right)
  \left(
  \begin{array}{c}
   \psi_A\\
   \psi_B
  \end{array}
  \right) = {\bf f}^{\ast}(\boldsymbol{\psi}; -k), \\
P^{-1}{\bf f}(P\boldsymbol{\psi};k) &=& \left(
  \begin{array}{cc}
   0 & 1 \\
   1 & 0
  \end{array}
  \right) \left(
  \begin{array}{cc}
   \kappa|\psi_B|^2 & a+be^{-ik} \\
   a+be^{ik} & \kappa|\psi_A|^2
  \end{array}
  \right)
  \left(
  \begin{array}{c}
   \psi_B\\
   \psi_A
  \end{array}
  \right) = \left(
  \begin{array}{c}
   (a+be^{ik}) \psi_B  + \kappa|\psi_A|^2 \psi_A \\
   (a+be^{-ik}) \psi_A + \kappa|\psi_B|^2 \psi_B
  \end{array}
  \right) = {\bf f}(\boldsymbol{\psi}; -k).
}

\subsection{\label{sec2}Calculation of bulk modes.}
Here, we analytically calculate the bulk eigenvalues and eigenvectors that satisfy Eq.~(6) in the main text. We first rewrite Eq.~(6) in the main text as
\begin{eqnarray}
(E-\kappa|\psi_A|^2)\psi_A &=& ce^{-i\theta} \psi_B, \label{eigeq1} \\
(E-\kappa|\psi_B|^2)\psi_B &=& ce^{i\theta} \psi_A, \label{eigeq2}
\end{eqnarray}
where $c$ and $\theta$ represent the absolute value and argument of $a+be^{ik}$, $c=|a+be^{ik}|$ and $\theta=\arg (a+be^{ik})$. Multiplying the left- (right-)hand side of the first equation by the right- (left-)hand side of the second one, we obtain
\begin{eqnarray}
(E-\kappa|\psi_A|^2)\psi_A^2 = e^{-2i\theta} (E-\kappa|\psi_B|^2)\psi_B^2. \label{reduced_eigeq}
\end{eqnarray}
We here assume that $a+be^{ik}$ is nonzero. If $a+be^{ik}$ is zero, the nonlinear eigenvalues and eigenvectors in Eq.~(6) become $E=\kappa w$, $(\psi_A,\psi_B)=(\sqrt{w},0),\ (0,\sqrt{w})$ and $E=\kappa w/2$, $(\psi_A,\psi_B)=(\sqrt{w/2},\sqrt{w/2})$ with $w$ being the squared norm of the eigenvector.

 To match the arguments of both hand sides of Eq.~\eqref{reduced_eigeq}, we need $2\arg(\psi_A) = 2\arg(\psi_B)-2\theta\  ({\rm mod}\,2\pi)$. Therefore, $\psi_{A,B}$ are written as 
\begin{eqnarray}
\psi_A = r_A e^{i\phi}, \label{psi_polarA}\\
\psi_B = \pm r_B e^{i(\phi+\theta)}, \label{psi_polarB}
\end{eqnarray}
by using real positive numbers $r_A$ and $r_B$ and representing the argument of $\psi_A$ as $\phi$. Substituting these representations into Eq.~\eqref{reduced_eigeq} and conducting a little algebraic calculation, we obtain
\begin{eqnarray}
(r_A^2-r_B^2)(\kappa(r_A^2+r_B^2)-E) = 0. \label{reduced_eigeq2}
\end{eqnarray}
Therefore, the nonlinear eigenvalue problem in Eq.~(6) in the main text has two types of solutions, which satisfy $r_A=r_B$ or $\kappa(r_A^2+r_B^2)=E$.

\begin{figure}[t]
  \includegraphics[width=160mm,bb=0 0 843 205,clip]{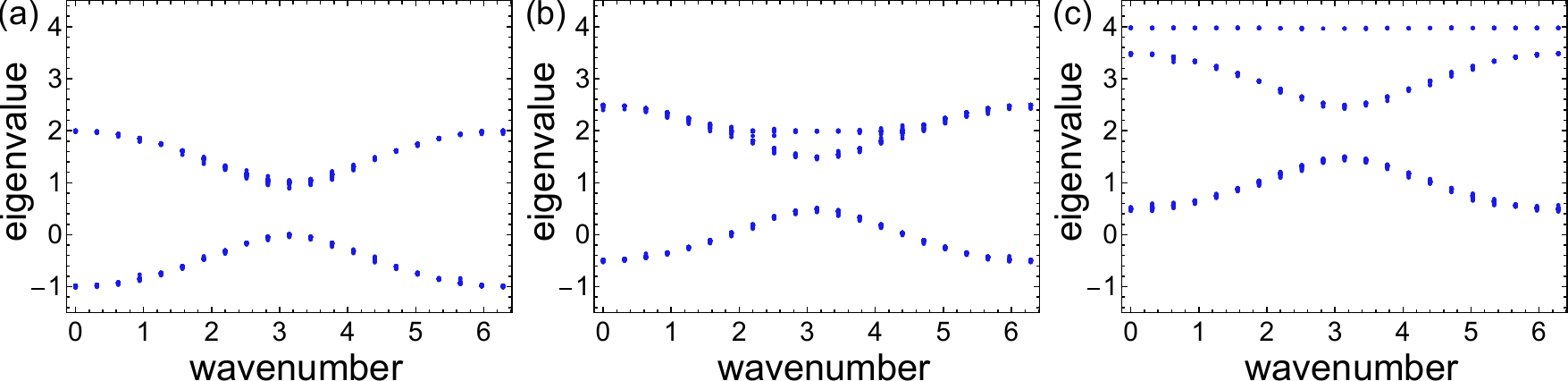}
\caption{\label{supplefig_bulk} Bulk band structures in the nonlinear SSH model. We use the parameters $a=-0.5$, $b=1$, $\kappa=1$, and fix the squared norm (a) $w=1$, (b) $w=2$, and (c) $w=4$. (a) At small amplitudes, the band structure is only uniformly shifted from the linear limit. $w=1$ is the critical point where nonlinear bulk modes without linear counterparts appear. (b) Above $w>1$, we obtain nonlinearity-induced flat bands (doubly degenerated) whose eigenvalues are $E=\kappa w$. (c) At larger amplitudes, the nonlinearity-induced flat bands are fully separated from the other bands.
}
\end{figure}

In the case of $r_A=r_B= \sqrt{w}$, substituting Eqs.~\eqref{psi_polarA} and \eqref{psi_polarB} into \eqref{eigeq1}, we obtain
\begin{eqnarray}
E-\kappa w = \pm c = \pm|a+be^{ik}|. \label{band1}
\end{eqnarray}
Therefore, the corresponding eigenvalues and eigenvectors are $E=\kappa w \pm |a+be^{ik}|$, $(\psi_A,\psi_B) = (\sqrt{w},\pm\sqrt{w}e^{i\theta})$.

In the case of $E=\kappa(r_A^2+r_B^2)=\kappa w$, substituting Eqs.~\eqref{psi_polarA}, \eqref{psi_polarB}, and this $w$-dependence of $E$ into \eqref{eigeq1}, we obtain
\begin{eqnarray}
r_A r_B = \pm \frac{c}{\kappa} = \pm \frac{|a+be^{ik}|}{\kappa}. \label{band2}
\end{eqnarray}
Therefore, when $\kappa>0$ and $\kappa w > 2|a+be^{ik}|$, we obtain the following eigenvectors with the eigenvalue $E=\kappa w$:
\begin{eqnarray}
   \left(
  \begin{array}{c}
   \psi_A\\
   \psi_B
  \end{array}
  \right) =   \left(
  \begin{array}{c}
   \sqrt{\left[\kappa w \pm \sqrt{(\kappa w)^2 - 4|a+be^{ik}|^2} \right] / (2\kappa)} \\
   e^{i\theta}\sqrt{\left[\kappa w \mp \sqrt{(\kappa w)^2 - 4|a+be^{ik}|^2} \right] / (2\kappa)} 
  \end{array}
  \right),
\label{nonlinear_bulk_modes}
\end{eqnarray}
where we must be careful for the argument of both hand sides of Eq.~\eqref{eigeq1} to omit the solutions with negative $r_A r_B$ in Eq.~\eqref{band2}.
We note that these eigenvectors disappear at small amplitudes $\kappa w < 2|a+be^{ik}|$, which indicates that these bulk modes can appear above certain strength of on-site nonlinearity. In the negative $\kappa$ case, the sign of $\psi_B$ is inverted, and the corresponding bulk modes appear at $-\kappa w > 2|a+be^{ik}|$.

One can also calculate the nonlinear Berry phase \cite{Zhou2022,Zhou2024} from the analytical expressions of the bulk bands. For any bulk bands, the nonlinear Berry phase becomes
\eqn{
\nu = \int_0^{2\pi} dk \frac{d\theta}{dk} = \pi\ ({\rm mod}\ 2\pi)
}
in the parameter region $|a|<|b|$. In the trivial case $|a|>|b|$, we obtain $\nu=0\ ({\rm mod}\ 2\pi)$.

We also conduct a numerical calculation of the bulk modes. We here regard Eq.~(6) in the main text as an algebraic equation of $\psi_A(k)$, $\psi_B(k)$, and $E$, and numerically solve it under the condition of $|\psi_A(k)|^2+|\psi_B(k)|^2=w$ by using the quasi-Newton method \cite{Sone2024a,Broyden1965}.  Supplementary Figure \ref{supplefig_bulk} shows the numerically obtained band structures at different $w$. These numerical results are consistent with the analytical calculations above.

\subsection{\label{sec4}Linear stability analysis of fixed points in the spatial dynamics.}
We here conduct the linear stability analysis of fixed points shown in Fig.~3 in the main text. We linearize the spatial dynamics in Eqs.~(8) and (9) by considering the deviations from the fixed points $\Psi_{A,B}^{\rm fix}$: $\delta\Psi_{A,B}(x) = \Psi_{A,B}(x) - \Psi_{A,B}^{\rm fix}$. Then, the linearized spatial dynamics becomes
\eqn{
   \left(
  \begin{array}{c}
   \delta\psi_A(x+1)\\
   \delta\psi_B(x+1)
  \end{array}
  \right) =  \left(
  \begin{array}{cc}
   -a/b & K_1 \\
   (-E + K_2)/b & (-b+EK_1 - K_1 K_2)/a
  \end{array}
  \right)
  \left(
  \begin{array}{c}
   \delta\psi_A(x)\\
   \delta\psi_B(x)
  \end{array}
  \right) \label{linear_stability}
}
with $K_1$ and $K_2$ being $K_1 = (E-3\kappa (\Psi_B^{\rm fix})^2)/b$ and $K_2 = 3\kappa(-a\Psi_A^{\rm fix}+E\Psi_B^{\rm fix}-\kappa (\Psi_B^{\rm fix})^3)^2/b^2$. We here focus on cases that all the parameters and $\Psi_{A,B}$ are real. We consider the parameters $a=-0.5$, $b=1$ and $\kappa=1$, and numerically diagonalize the above matrix at $(\Psi_A^{\rm fix},\Psi_B^{\rm fix})=(0,0)$, $(\sqrt{2(E-a-b)/\kappa},\sqrt{2(E-a-b)/\kappa})$, and $(r,|a+b|/(\kappa r))$ with $r$ being $r=\sqrt{\left[E + \sqrt{E^2 - 4|a+b|^2} \right] / (2\kappa)}$, which correspond to the bulk eigenvectors with wavenumber$k=0$ (Eqs. (7) and (8) in the main text).

\begin{figure}[t]
  \includegraphics[width=80mm,bb=0 0 280 210,clip]{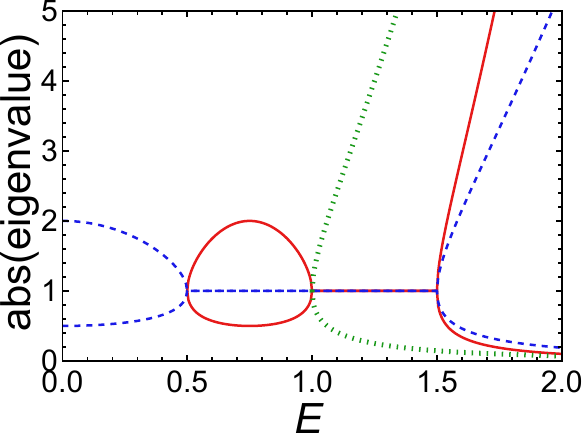}
\caption{\label{supplefig_linear_stability} Linear stability analysis of fixed points in the spatial dynamics.  We calculate the eigenvalues of the matrix in Supplementary Eq.~\eqref{linear_stability} and show their absolute values at different $E$. We use the parameters $a=-0.5$, $b=1$, $\kappa=1$. We consider the linearization around fixed points at $(\Psi_A^{\rm fix},\Psi_B^{\rm fix})=(0,0)$ (the blue dashed curve), $(\Psi_A^{\rm fix},\Psi_B^{\rm fix})=(\sqrt{2(E-a-b)/\kappa},\sqrt{2(E-a-b)/\kappa})$ (the red curve), and $(\Psi_A^{\rm fix},\Psi_B^{\rm fix})=(\sqrt{\left[E + \sqrt{E^2 - 4|a+b|^2} \right] / (2\kappa)}, \sqrt{\left[E - \sqrt{E^2 - 4|a+b|^2} \right] / (2\kappa)})$ (the green dot curve). We plot the red (green dot) curve only in the range of $E>|a+b|$ ($E>2|a+b|$) because the corresponding fixed points disappear below those parameter regions. When the absolute values of eigenvalues are one, the fixed point is linearly marginal and otherwise, it is a saddle point. At both the nonlinearity-induced transition points, $E=|a+b|$ and $E=2|a+b|$, the linear stabilities of fixed points are changed, which indicates their bifurcations.
}
\end{figure}

Supplementary Figure \ref{supplefig_linear_stability} shows the absolute values of the eigenvalues of the matrix in Supplementary Eq.~\eqref{linear_stability}. If the absolute values are not one, the fixed point is a saddle point at that parameter. Then, there is a one-dimensional stable manifold of the saddle point, along which a nonlinear edge mode is spatially changed. At $E=|a+b|$ and $E=2|a+b|$, the linear stabilities of fixed points are altered, and thus bifurcations occur. Such bifurcations induce the transition of nonlinear edge modes. In fact, after the bifurcation, the absolute values of eigenvalues in Eq.~\eqref{linear_stability} are one, which means that the fixed points are centers. Then, there are many periodic solutions around them, and they have no stable manifolds (instead they have center manifolds). Thus, edge modes changing along their stable manifold disappear by the bifurcation. We note that when centers appear, periodic solutions around them seem to correspond to other bulk modes with different wavenumbers. In fact, for $0.5<E<1.5$, the origin is a center and the nonlinear SSH model also has bulk modes. We remain the comprehensive understanding of the relationship between linear stability and the existence of bulk modes as a future perspective. We also note that for $E>1.5$, all the fixed points considered in Supplementary Fig. \ref{supplefig_linear_stability} become saddle points, while this does not indicate the existence of topological localized modes corresponding to all of these fixed points because the stable manifold may not cross the $\Psi_B=0$ axis and the localization properties are also unpredictable. We again emphasize that Supplementary Fig. \ref{supplefig_linear_stability} indicates the bifurcations of fixed points at $E=0.5$ and $1.0$, which induces the nonlinearity-induced transition.

We note that the determinant of the matrix in Supplementary Eq.~\eqref{linear_stability} is one independently of the values of $E$ and $\Psi_{A,B}^{\rm fix}$. This is consistent with a transfer matrix of a linear Hermitian system. In the parameter region where the linearized matrix around $(0,0)$ has eigenvalues $|\lambda|=1$, there are corresponding bulk bands, which is also reminiscent of the linear cases. We remain the full understanding of the relationship between nonlinear bands and eigenvalues of linearized matrices to future studies.

\subsection{\label{sec5}Bifurcation corresponding to the second nonlinearity-induced transition at $E=2|b-a|$.}
We have found two nonlinearity-induced transition points $E=a+b$ and $E=2(a+b)$ in the nonlinear SSH model as shown in Fig.~2 in the main text. We here discuss that the second transition point also accompanies the bifurcation of the spatial dynamics (Eqs.~(8) and (9) in the main text) as the first one does. Such a bifurcation is related to the emergence of the nonlinear bulk modes in Supplementary Eq.~\eqref{nonlinear_bulk_modes}.

The nonlinear edge modes at $a+b<E<2(a+b)$ corresponds to the converging orbit to a fixed point at $(\psi_A,\psi_B) = (\pm \sqrt{w/2},\pm \sqrt{w/2})$ as discussed in the main text. However, this fixed point becomes linearly marginal at $E>2(a+b)$ 
(see the previous section). At this transition point of the linear stability, one can find the bifurcation of the fixed point. In fact, fixed points appear at $(\psi_A,\psi_B) = (\sqrt{\left[E \pm \sqrt{E^2 - 4|a+b|^2} \right] / (2\kappa)},\sqrt{\left[E \mp \sqrt{E^2 - 4|a+b|^2} \right] / (2\kappa)})$, which corrresponds to the nonlinearity-induced bulk modes.

\begin{figure}[t]
  \includegraphics[width=80mm,bb=0 0 305 183,clip]{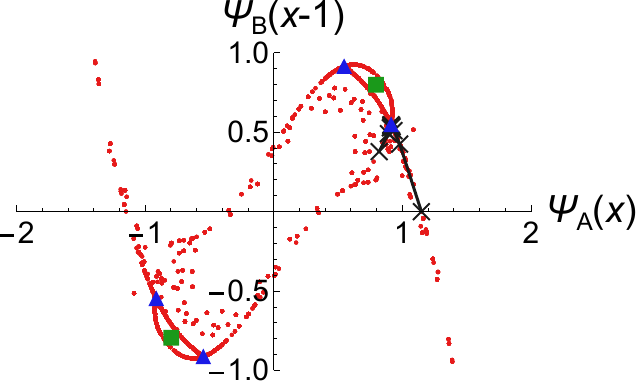}
\caption{\label{supplefig_bifurcation} Stable manifolds demonstrating the bifurcation at the second nonlinearity-induced transition point. At the second transition point ($E=2|a+b|$), the fixed points represented by the green squares bifurcate, and new fixed points appear as represented by the blue triangles. The red dots represent the numerically estimated stable manifolds of the fixed points at the blue triangles. The black crosses and polyline show the spatial dynamics of a topological gapless mode in Fig.~2 in the main text. We can confirm that the values of the wave function change along the stable manifold.
}
\end{figure}

We here numerically demonstrate such a bifurcation by calculating the stable manifolds of the newly emerging fixed points. In Supplementary Fig.~\ref{supplefig_bifurcation}, we plot the stable manifolds as in Fig.~3 in the main text. We can confirm the split of fixed points and that the topological edge modes correspond to the orbit approaching the new fixed point. Since the stable manifold surrounds the fixed points at $(\sqrt{2(E-a-b)/\kappa},\sqrt{2(E-a-b)/\kappa})$, the spatial dynamics of edge modes does not converge to those original fixed points, and thus the bifurcation induces the nonlinearity-induced transition of topological edge modes.

We note that the stable manifolds of fixed points form curves as in Fig.~3 in the main text, even though the dynamical system is discrete. This is natural as long as the nonlinear effect is perturbative, because the stable manifold is approximated by the stable linear subspace, $\{(\Psi_A^{\rm fix}+\Delta_A,\Psi_B^{\rm fix}+\Delta_B)\,|\,(\Delta_A,\Delta_B)=t \mathbf{v}\}$, where $\mathbf{v}$ is the eigenvector of the linearized matrix \eqref{linear_stability} with the absolute value of the eigenvalue less than one. The perturbative effect is approximated by the lower-order expansion $(\Delta_A,\Delta_B)=t \mathbf{v} + t^2 \mathbf{v}_2 + t^3 \mathbf{v}_3 + \cdots$, which still represents a curve. It is also noteworthy that some of the stable manifolds are the unstable manifolds of other fixed points, i.e., there are heteroclinic orbits. Since nonlinearity-induced topological localized modes have bulk values close to the fixed points, their values at the left (right) edge are close to the stable (unstable) manifolds (this is because of the characteristic properties of the stable (resp. unstable) manifolds that they are a set of points to which the dynamics converge at $x\rightarrow -\infty$ (resp. $x\rightarrow \infty$)).

\subsection{\label{sec6}Details of numerical calculation in Fig. 3.}
In Fig.~3 in the main text, we plot the numerically estimated stable manifolds in the spatial dynamics in Eqs.~(8) and (9). We here utilize the fact that the stable manifold is a set of points to which the inverse dynamics converge at $x\rightarrow -\infty$. Since the spatial dynamics in Eqs.~(8) and (9) is derived from the nonlinear eigenvalue problem (Eqs.~(4) and (5) in the main text), one can straightforwardly derive its inverse dynamics as 
\begin{eqnarray}
&{}& \Psi_B(x-1) = \frac{-a\Psi_B(x)-\kappa|\Psi_A(x)|^2\Psi_A(x)+E\Psi_A(x)}{b}, \label{transfer_inv1}\\
&{}& \Psi_A(x-1) = \frac{-b\Psi_A(x)-\kappa|\Psi_B(x-1)|^2\Psi_B(x-1)+E\Psi_B(x-1)}{a},\label{transfer_inv2}
\end{eqnarray}
by rewriting the nonlinear eigenvalue problem into the spatial dynamics in the inverse direction. In the numerical calculation in FIg.~3, we set the initial condition as the neighbor points of the fixed points, $(\Psi_{A}(0),\Psi_{B}(0))=(\Psi_A^{\rm fix} + \Delta r \cos \theta, \Psi_A^{\rm fix} + \Delta r \sin \theta)$ with $(\Psi_A^{\rm fix},\Psi_B^{\rm fix})$ representing the fixed point. We use several $\Delta r$ and $\theta$: $\Delta r = 10^{-4} n$, $n=1,\ldots,10$ and $\theta = m\pi/ 6$, $m=1,\ldots,12$. Finally, we plot the values of $(\Psi_{A}(x),\Psi_{B}(x))$ at $x<-10$. We have also used the same numerical technique in Supplementary Fig.~\ref{supplefig_bifurcation}, while we set $\Delta r$ and $\theta$ as $\Delta r$ and $\Delta r = 0.5\times10^{-15} n$, $n=1,\ldots,30$ and $\theta = m\pi/ 12$, $m=1,\ldots,24$ because the nonlinear spatial dynamics becomes more sensitive to the initial condition than in Fig.~3.

\subsection{\label{sec7}Details of numerical simulation in Fig. 4.}
In Fig.~4 in the main text, we numerically simulate the dynamics in Eqs.~(1) and (2) by using the fourth-order Runge-Kutta method. We set the time step as $\Delta t=0.0005$ and use the parameters $a=-0.5$, $b=1$, $\kappa=1.0$. The initial condition of the simulation is $\Psi_A(1) = \sqrt{0.1}$ and $\Psi_A(x\geq 2) = \Psi_B(x\geq 1) = 0$. We consider the system size $L=1000$, while we only plot the amplitude of the nonlinear wave in the range of $1\leq x \leq 25$ in FIg.~4 in the main text. We need a large system to observe the nonlinearity-induced transition because if a propagating wave reflecting at the right edge comes back to the left edge, the localized wave is perturbed and destroyed by their nonlinear interaction.

At the first stage of the simulation, we introduce no gains and wait until $t=t_0=200$ when the dynamics is relaxed around the left edge and forms a localized state. Then, we introduce a uniform gain, which adds a diagonal term $i c(t) I$ into Eq.~(6) in the main text and whose strength depends on the time as $c(t) = \left[ 2(t-t_0)+1/c_0 \right]^{-1}$, $c_0=0.2$. This decaying gain rate is necessary to suppress the speed of deformation of nonlinearity and realize approximately adiabatic deformation of the nonlinear system. In fact, if we consider an isolated site for simplicity, the time dependence of its amplitude becomes $|\Psi|^2 = (\sqrt{2(t-t_0)+1/c_0}-\sqrt{1/c_0}+\psi_0)^2 = \mathcal{O}(t)$. At the final stage of the simulation, we switch off the uniform gain at $t=t_1=300$ and simulate the free time evolution of the nonlinear SSH model.

\end{document}